\documentclass[aps,preprint,nofootinbib,superscriptaddress,prc]{revtex4}
\usepackage{amssymb}

\usepackage{epsfig}
\usepackage[bookmarksnumbered,bookmarksopen,colorlinks,citecolor=blue,linkcolor=blue]{hyperref}
\begin{document}

\title{Determination of the nucleon-nucleon interaction in the ImQMD model by nuclear reaction at the Fermi energy region}

\author{Cheng Li }

 \affiliation{Department of Physics,
Guangxi Normal University, Guilin 541004, P. R. China}
 \affiliation{School of Physics and Electrical Engineering, Anyang Normal
University, Anyang 455000, P. R. China}

\author{Junlong Tian }
\email{tianjunlong@gmail.com}  \affiliation{School of Physics and
Electrical Engineering, Anyang Normal University, Anyang 455000, P.
R. China}

\author{Yujiao Qin }
\affiliation{Department of Physics, Guangxi Normal University,
Guilin 541004, P. R. China}

\author{Jingjing Li }
  \affiliation{School of Physics and
Electrical Engineering, Anyang Normal University, Anyang 455000, P.
R. China}

\author{Ning Wang }
\email{wangning@gxnu.edu.cn} \affiliation{Department of Physics,
Guangxi Normal University, Guilin 541004, P. R. China}

\begin{abstract}
The nucleon-nucleon interaction is investigated by using the ImQMD model with the three sets of parameters IQ1, IQ2 and IQ3 in which the corresponding incompressibility coefficients of nuclear matter are different. The charge distribution of fragments for various reaction systems are calculated at different incident energies. The parameters strongly affect the charge distribution below the threshold energy of nuclear multifragmentation. The fragment multiplicity spectrum for $^{238}$U+$^{197}$Au at 15 AMeV and the charge distribution for $^{129}$Xe+$^{120}$Sn at 32 and 45 AMeV, and $^{197}$Au+$^{197}$Au at 35 AMeV are reproduced by the ImQMD model with the set of parameter IQ3. It is concluded that charge distribution of the fragments and the fragment multiplicity spectrum are good observables for studying N-N interaction, the Fermi energy region is a sensitive energy region to explore the N-N interaction, and IQ3 is a suitable set of parameters for the ImQMD model.
\end{abstract}

\maketitle

\begin{center}
\textbf{I. INTRODUCTION}
\end{center}

$~~~$The nucleon-nucleon (N-N) interaction is the most fundamental problem in nuclear physics. It is related to many nuclear properties and nuclear reactions mechanism, for example, the binding energies, incompressibility coefficient, nuclear structure, fusion-fission reactions and so on. Therefore, knowing the N-N interaction potential form is of great significance for us to explain the nature of nucleus and study nuclear reaction mechanism. The Skyrme force is an effective N-N interaction with various parameterizations that have been proposed to the G-matrix for nuclear Hartree-Fock calculations reproducing the basic nuclear structure. Since the first work of Vautherin and Brink\cite{1}, who performed fully microscopic self-consistent mean-field Hartree-Fock calculations with the Skyrme type effective nucleon-nucleon interaction\cite{2,3,4}, many different parameterizations of the Skyrme interaction have been proposed and have been reproduced the data of the nuclear masses, radii and other physical quantities, such as SkP, SkM$^*$, Sly1-7\cite{5,6,7}. The improved quantum molecular dynamic (ImQMD) model adopts the Skyrme type effective interaction and is successfully used for intermediate-energy heavy-ion collisions and heavy-ion reactions at energies near the Coulomb barrier\cite{8,9,10,11}. By combining the known Skyrme forces and the experimental data for fusion reactions and heavy-ion collisions at intermediate energies, three different parameter sets IQ1, IQ2 and IQ3 were proposed\cite{9,12,13}. The different ImQMD parameters mean different N-N interaction, and they are associated with different nuclear equations of state (EOS).\\
$~~~~~~$In low energy heavy-ion collision, the reaction system has a low excitation energy, one observes the emission of light particles plus an evaporation residue for light systems or fission caused by Coulomb repulsion for heavy systems. As the incident energy increases to a threshold energy, the excitation energy of the reaction system will reach maximum limit. It implies that the nuclear reaction will enter the multifragmentation process. In order to investigate N-N interaction, we will study the charge distribution of fragments for various reaction system at different incident energies and the fragment multiplicity spectrum for $^{238}$U+$^{197}$Au at 15 AMeV with the ImQMD model by adopting different parameters. The structure of this paper is as follows: In Sec. {\bf II}, relation of EOS and wave-packet width is briefly introduced. In Sec. {\bf III}, three sets of ImQMD parameters IQ1, IQ2 and IQ3 will be employed to calculate charge distribution for $^{40}$Ca+$^{40}$Ca with incident energy from 10 to 45 AMeV, $^{129}$Xe+$^{120}$Sn at 32 and 45 AMeV, $^{197}$Au+$^{197}$Au at 35 and 50 AMeV, and fragment multiplicity spectrumat at 15 AMeV for $^{238}$U+$^{197}$Au. Finally, a summary is given in Sec. {\bf IV}.

\begin{center}
\textbf{II. EOS AND WAVE-PACKET WIDTH}
\end{center}

The equation of state (EOS) of nuclear matter plays an important role in studying of nuclear properties, heavy-ion collisions, neutron stars and supernova. The EOS depends on the interactions and properties of the particles in the matter. It describes how the state of the matter changes under different conditions. For cold nuclear matter, the EOS is usually defined as the binding energy per nucleon as a function $E/A=\varepsilon(\rho,T,\delta)$ of the density $\rho$, temperature $T$ and and isospin asymmetry $\delta=(\rho_n-\rho_p)/\rho$. For the symmetric nuclear matter $\delta=0$ \cite{14}. Empirical values of the nuclear matter EOS such as energy per nucleon, incompressibility and saturation density of symmetric nuclear matter at $T=0$ MeV suggest that $\varepsilon_0\approx-16$ MeV, $K_\infty\approx230$ MeV and $\rho_0\approx0.16$ fm$^{-3}$. However, the uncertainty of nuclear equation of state still causes some difficulties for an unambiguous determination of the model parameters. In the ImQMD model, the equation of state of the symmetric nuclear matter can be expressed as \cite{12,15}

\begin{equation}
\frac{E(\rho)}{A} =\xi c_k\rho^{2/3}+\frac{1}{2}\alpha\frac{\rho}{\rho_0}
+\beta\frac{\rho^\gamma}{(\gamma+1)\rho_0^\gamma}
+g_\tau\frac{\rho^\eta}{\rho_0^\eta}. \label{aba:app1}
\end{equation}Where $\xi=c_0/c_k$, $c_k=\frac{3}{5}\frac{\hbar^2}{2m}(\frac{3\pi^2}{2})^{2/3}=75.0 MeV fm^{-2}$. The coefficient $c_0$ can be determined by the kinetic energies of nuclei at their ground state \cite{15}. The density distribution function $\rho$ of a system can be read \begin{equation}
\rho(r)=\sum_{i}\frac{1}{(2\pi\sigma_r)^{3/2}}\exp[-\frac{(r-r_i)^2}{2{\sigma_r}^2}]. \label{aba:app1}
\end{equation}Fig. 1 shows the energy per nucleon of symmetric nuclear matter as a function of $\rho/\rho_0$. It is seen that the EOS with the IQ3 is clearly harder than that with the IQ1 and IQ2 in region $\rho/\rho_0>1$ with the increase of the density. For the parameter set IQ3, the incompressibility coefficient is 226 MeV and much higher than the corresponding values for IQ1 (165 MeV) and IQ2 (195 MeV). In heavy-ion collision process, the projectile and target firstly contacted, compressed and then expanded. The soft nuclear matter have a higher compression density and more violent expansion process. While for the finite nuclear system, the expansion process will be influenced by the nuclear surface energy, Coulomb energy and wave-packet width, etc.\\
$~~~~~~$In the ImQMD model, each nucleon is described by a coherent state of Gaussian wave packet. The system-size-dependent wave-packet width in coordinate space is given by the  formula: \begin{equation}
\sigma^{n}_r=\sigma_0+\sigma_1A_n^{1/3},  \; \; n=\{p,t\}.
\end{equation}
Here, $\sigma^{p}_r$ ($\sigma^{t}_r$) denotes the wave-packet width
for the nucleons which belong to the projectile (target). $A_p$ and
$A_t$ denote the mass number of the projectile and target,
respectively. For symmetric reaction systems,
$\sigma_r=\sigma^{p}_r=\sigma^{t}_r$. The parameter $\sigma_0$ and $\sigma_1$ have been listed in Table 1. The wave-packet width is useful for exploring the influence of the interaction range of nucleons and the finite-size effect of nuclei. In Fig. 2, we show the wave-packet width of nucleon as a function of system-size with the three sets of parameters. One can see from the figure that there is a large difference with the different system. The wave-packet width given by the IQ1 and IQ2 obviously is higher than the IQ3 with increasing mass number. The three sets of parameters are listed in Table 1.

\begin{figure}\center
\psfig{file=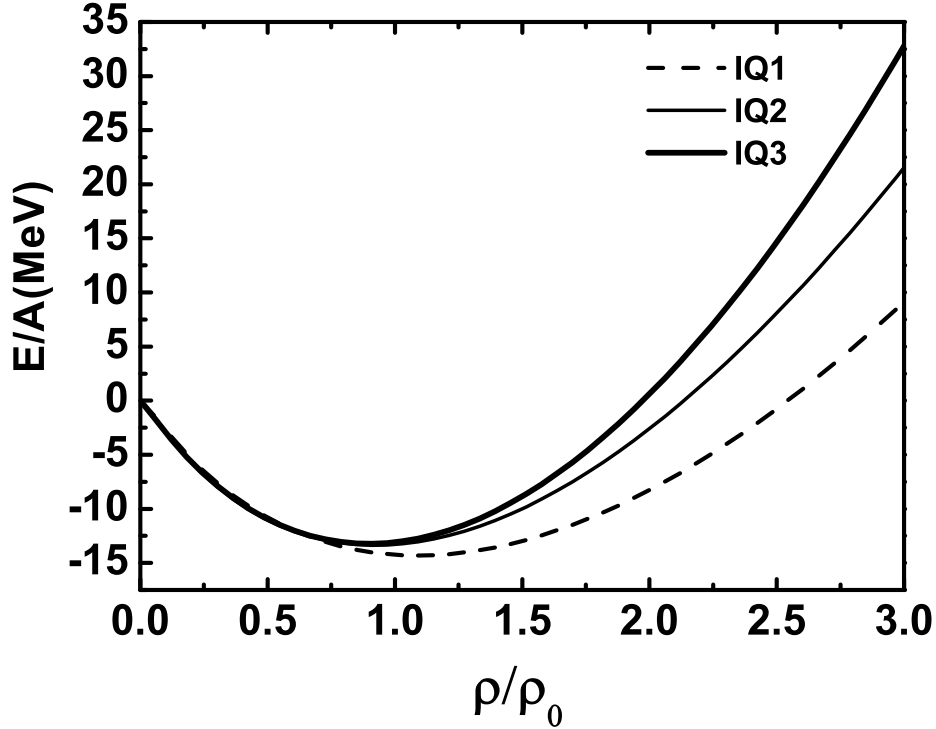,width= 0.75\textwidth} \caption{The energy per nucleon of symmetric nuclear matter for different parameters.} \label{aba:Fig1}
\end{figure}

\begin{figure}\center
\psfig{file=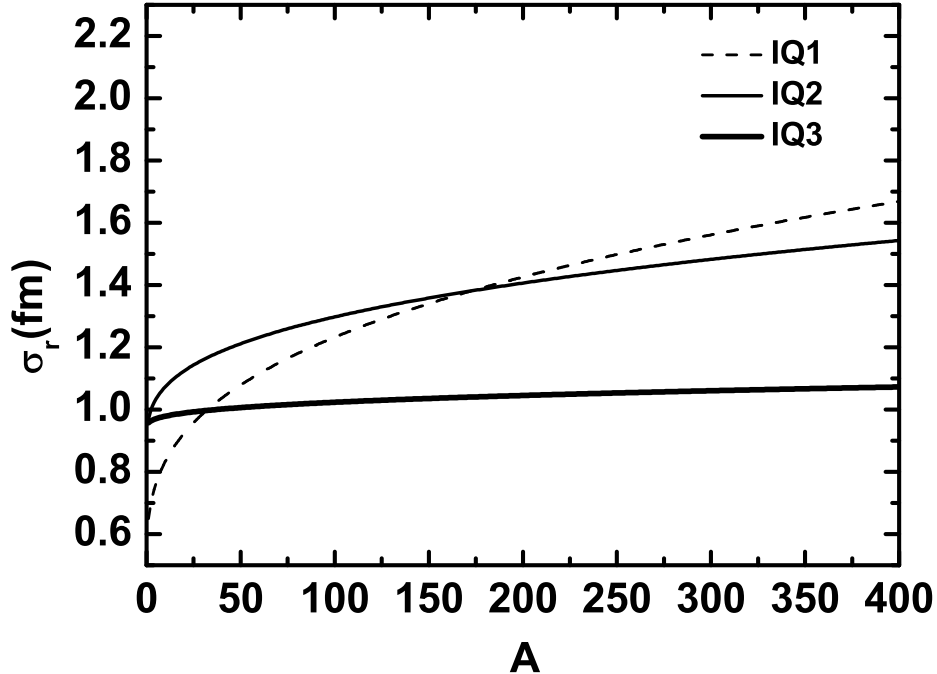,width= 0.75\textwidth} \caption{The wave-packet width as a function of system-size with the three sets of parameters.} \label{aba:Fig2}
\end{figure}

\begin{table}
\tabcolsep=1pt \caption{The model parameters.}
{\begin{tabular}{@{}cccccccccccc@{}}

\hline\hline
  $$ & $\alpha$ & $\beta$ & $\gamma$ & $g_{0}$ & $g_{\tau}$ & $\eta$ & $C_{S}$ & $\kappa_{s}$ & $\rho_{0}$ & $\sigma_0$ & $\sigma_1$\\

$$ & $(MeV)$ & $(MeV)$ & $$ & $(MeVfm^{2})$ & $(MeV)$ & $$ & $(MeV)$ & $(fm^{2})$ & $(fm^{-3})$ & $(fm)$ & $(fm)$\\
\hline
  $IQ1$ & $-310$ & $258$ & $7/6$ & $19.8$ & $9.5$ & $2/3$ & $32.0$ & $0.08$ & $0.165$ & $0.49$ & $0.16$ \\

  $IQ2$ & $-356$ & $303$ & $7/6$ & $7.0$ & $12.5$ & $2/3$ & $32.0$ & $0.08$ & $0.165$ & $0.88$ & $0.09$\\

  $IQ3$ & $-207$ & $138$ & $7/6$ & $18.0$ & $14.0$ & $5/3$ & $32.0$ & $0.08$ & $0.165$ & $0.94$ & $0.018$\\
\hline\hline
\end{tabular}}

\end{table}

\begin{center}
\textbf{III. RESULTS}
\end{center}

Based on the ImQMD model, the charge distribution of fragments for different reaction systems will be employed to explore the N-N interaction. We present the charge distribution of fragments for $^{40}$Ca+$^{40}$Ca  at incident energy of 35 AMeV in Fig. 3. The solid triangles, solid squares and solid stars denote the results with the IQ1, IQ2 and IQ3, respectively. The open circles denote the experimental data\cite{17}. Here we create 500 events for central collisions and for each event we simulate the whole collision process until t=3000 fm/c without combining statistical model. From the Fig. 3a-3c, one sees that the experimental data can be reproduced very well with the ImQMD calculations with three sets of parameters. However, their nuclear equations of state and the wave-packet width are different. Why do they have a similar charge distribution?

\begin{figure}\center
\psfig{file=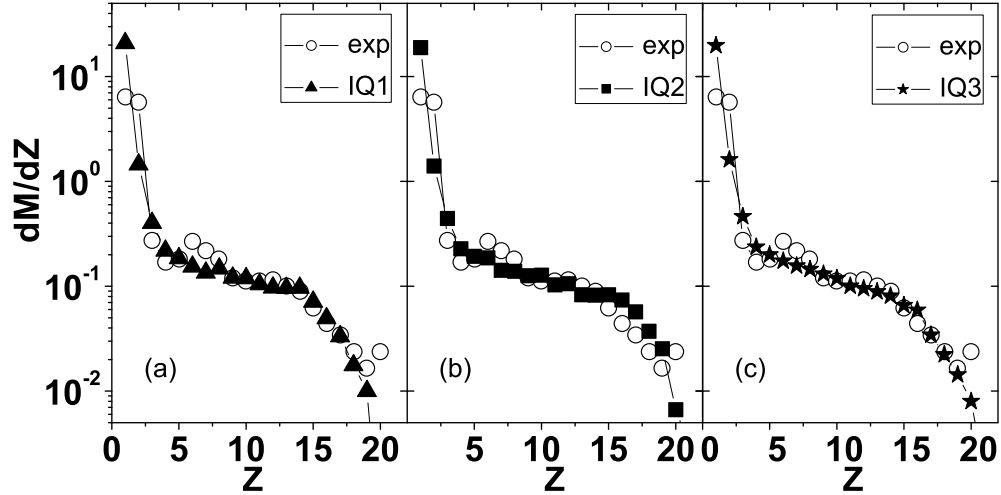,width= 0.8\textwidth} \caption{The charge distribution of fragments from three sets of parameters for $^{40}$Ca+$^{40}$Ca at incident energy of 35 AMeV.} \label{aba:Fig3}
\end{figure}

For further investigating the effect of N-N interaction on the heavy-ion collisions, we studied the charge distribution of fragments for different reaction system at different incident energies. In Fig. 4 we show the charge distribution of fragments calculated by the ImQMD model with three sets of parameters for $^{40}$Ca+$^{40}$Ca at incident energies from 10 to 45 AMeV. One sees that the results from IQ1, IQ2 and IQ3 for the charge distribution of fragments are relatively close to each other at the energy region 35-45 AMeV, but there exist significant differences at the energy region 10-30 AMeV. One can also see that all of the peak from the IQ1, IQ2 and IQ3 become gradually lower with increasing incident energies at the energy region 10-30 AMeV. In this region, it is the transition region of the fusion reaction to the multifragmentation for $^{40}$Ca+$^{40}$Ca reaction system. The fusion evaporation of reaction system is strongly affected by the ImQMD parameters. Above 30 AMeV, the nuclear reaction turn to the multifragmentation, the distribution of fragments is not sensitive to the ImQMD parameters.

\begin{figure}\center
\psfig{file=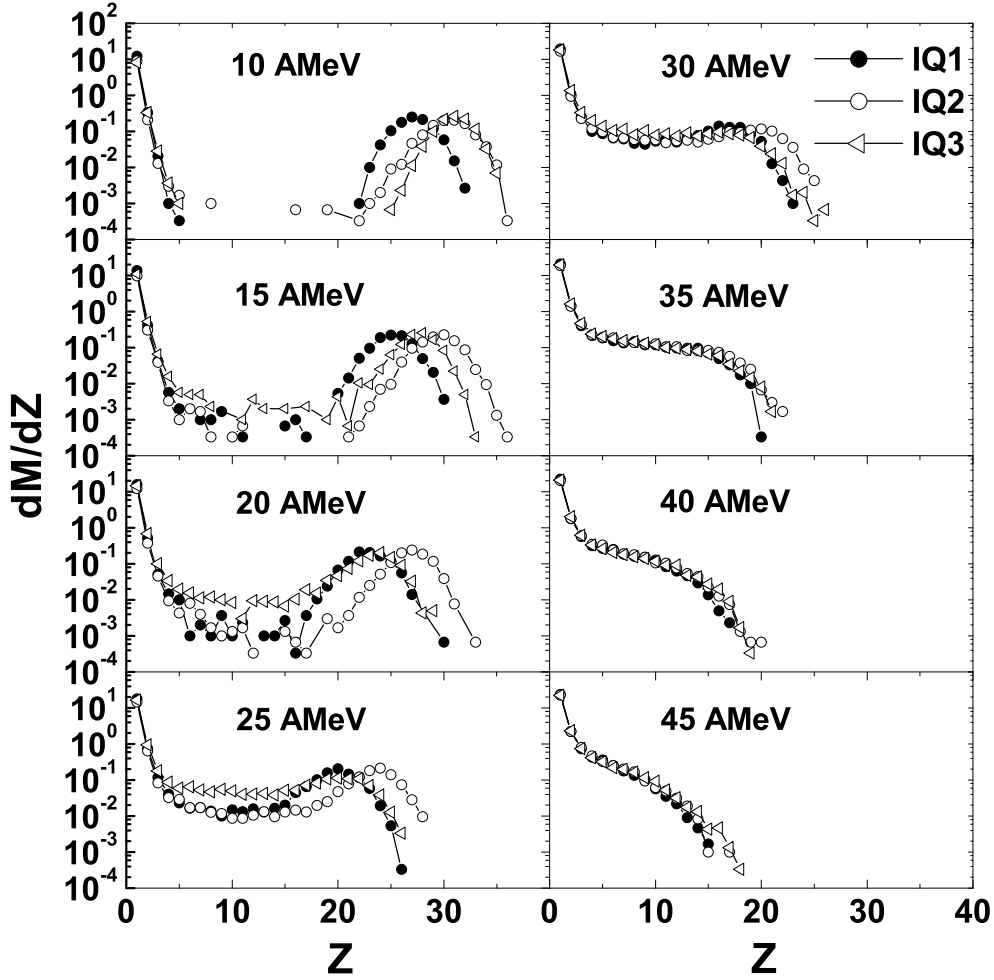,width= 0.8\textwidth} \caption{The charge distribution of fragments from three sets of parameters for $^{40}$Ca+$^{40}$Ca at incident energies from 10 to 45 AMeV. The solid circles, open circles and open triangles denote the results with the IQ1, IQ2 and IQ3, respectively.} \label{aba:Fig4}
\end{figure}

Fig. 5 shows the charge distribution of fragments for $^{129}$Xe+$^{120}$Sn at 32 and 45 AMeV \cite{18}, $^{197}$Au+$^{197}$Au at 35 and 50 AMeV\cite{19}, which are calculated by the ImQMD model using three sets of parameters. One can see that the charge distribution of fragments at 45 AMeV for $^{129}$Xe+$^{120}$Sn and 50 AMeV for $^{197}$Au+$^{197}$Au are also relatively close to each other calculated by three sets of parameters. While there still exist significant differences at the lower incident energy of 32 AMeV for $^{129}$Xe+$^{120}$Sn reaction and 35 AMeV for $^{197}$Au+$^{197}$Au reaction. Compare $^{129}$Xe+$^{120}$Sn and $^{197}$Au+$^{197}$Au reaction system with $^{40}$Ca+$^{40}$Ca, we find that the sensitive energy region for the ImQMD parameters is depend on the reaction systems. The beginning of insensitive energy region for heavy system is higher than the light system. It is well known, the mean field plays a dominant role at the low energy region. However, the nucleon-nucleon collision becomes more and more important with increasing energy. In the transition region, the mean field and the nucleon-nucleon collision work together, so it is a sensitive energy region for the N-N interaction. By comparing with the experimental results, we find that the the experimental data can be always reproduced very well by the IQ3. The IQ3 is a suitable set of parameters in the ImQMD model for heavy-ion collisions.

\begin{figure}\center
\psfig{file=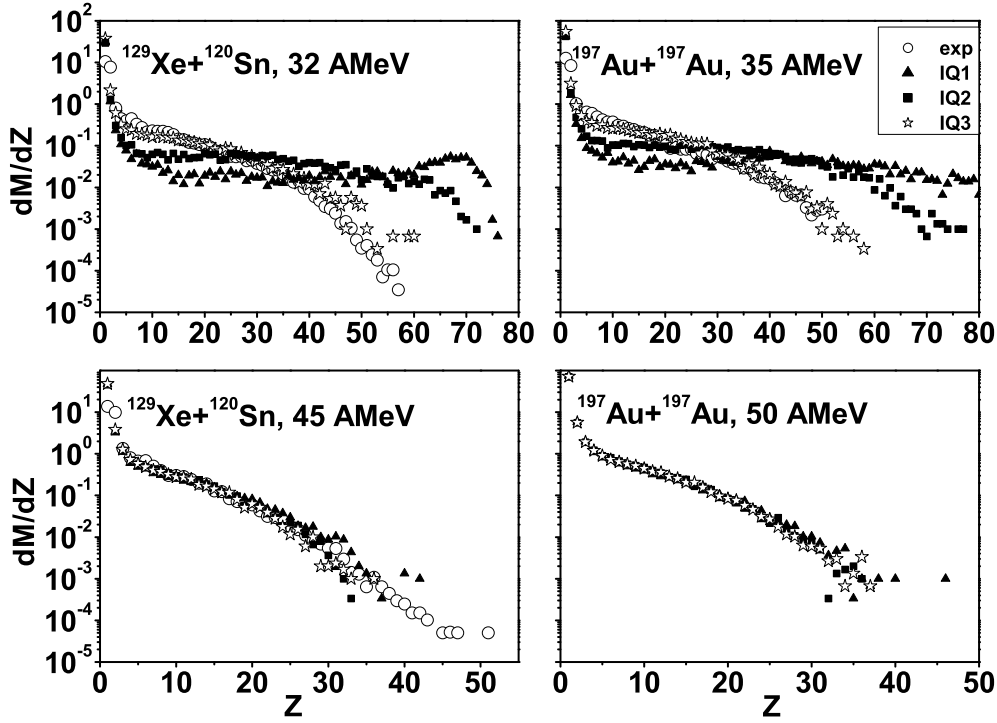,width=0.8\textwidth} \caption{The charge distribution of fragments for $^{129}$Xe+$^{120}$Sn at 32 and 45 AMeV, $^{197}$Au+$^{197}$Au at 35 and 50 AMeV. The solid triangles, solid squares, open stars and open circles denote the results with the IQ1, IQ2, IQ3 and the experimental data, respectively. The $^{197}$Au+$^{197}$Au at 50 AMeV does not give the experimental data.} \label{aba:Fig5}
\end{figure}

 In order to further test the above conclusion, we calculated fragment multiplicity spectrum of $^{238}$U+$^{197}$Au reaction by using three sets of parameters. In Fig. 6, we show the comparison of calculation results with the experimental data for fragment multiplicity spectrum of $^{238}$U+$^{197}$Au reaction at 15 AMeV. The open squares, open triangles, solid stars and open circles denote the results with IQ1, IQ2, IQ3 and the experimental data\cite{20}, respectively. We created 4000 events from central to peripheral collisions and counted the number of fragments in each reaction except in the case where the charge $Z<8$ as that did in the experiment. The experimental data shows that two-body events exhaust only about 5\% and the three- and four-body events exhaust approximately 83\% of the total reaction events. From Fig. 6, one can see that the ImQMD calculation result reproduce the experimental multiplicity spectrum reasonably well with the IQ3 rather than the IQ1 and the IQ2. It implies that the IQ3 is reasonable in the ImQMD model for heavy-ion collisions.

 \begin{figure}\center
\psfig{file=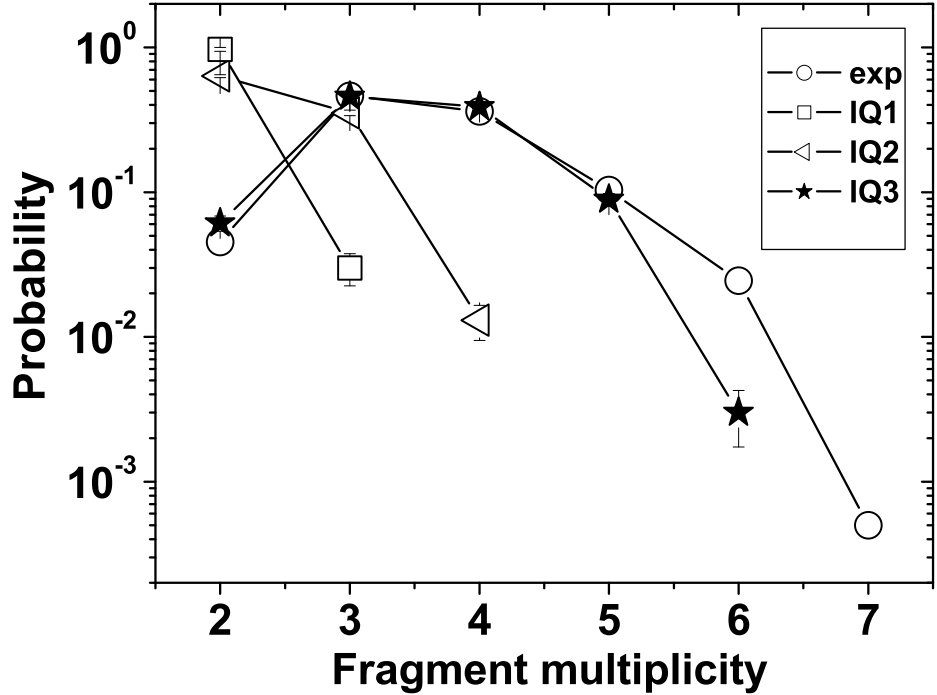,width=0.75\textwidth} \caption{The fragment multiplicity spectrum of $^{238}$U+$^{197}$Au reaction at 15 AMeV.} \label{aba:Fig6}
\end{figure}

\begin{center}
\textbf{IV. CONCLUSION}
\end{center}

The N-N interaction has been investigated by using the ImQMD model with the three sets of parameters IQ1, IQ2 and IQ3. The charge distribution of fragments at different incident energies were calculated based on the three sets of parameters. The calculation results demonstrate that the charge distribution and the fragment multiplicity spectrum are good observables and the Fermi energy region is a sensitive energy region for probing the N-N interaction. The charge distribution of fragments below the threshold energy of nuclear multifragmentation are very sensitive to the ImQMD parameters. By comparing the charge distribution of fragments and fragment multiplicity spectrum with the experimental data, we find that the IQ3 is a suitable set of parameters for the ImQMD model.

\begin{center}
\textbf{ACKNOWLEDGEMENTS}
\end{center}

One of the authors (TIAN Jun-Long) is grateful to Prof. ZHANG Ying-Xun
for fruitful discussions. This work was supported by the National
Natural Science Foundation of China (Nos. 11005003, 10975095 ,11275052 and
11005002), the Natural Science Foundation of He'nan Educational
Committee (Nos. 2011A140001,  2011GGTS-147) and innovation fund of undergraduate at Anyang Normal University (ASCX/2012-Z28).

\end{document}